# GPS Spoofing Attacks and Pilot Responses Using a Flight Simulator Environment

Mathilde Durieux
*École de l'air et de l'espace*
Chemin St Jean
13300 Salon-de-Provence, France
mathilde.durieux@ecole-air.fr

Kayla D. Taylor
*Department of Electrical Engineering and Computer Science*
Embry-Riddle Aeronautical University
1 Aerospace Blvd
Daytona Beach, FL, USA 32114
taylok33@my.erau.edu

Laxima Niure Kandel
*Department of Electrical Engineering and Computer Science*
Embry-Riddle Aeronautical University
1 Aerospace Blvd
Daytona Beach, FL, USA 32114
niurekal@erau.edu

Deepti Gupta
*Subhani Department of Computer Information Systems*
Texas A&M University-Central Texas
1001 Leadership Pl
Killeen, TX 76549
d.gupta@tamuct.edu

*Abstract*—Global Positioning System (GPS) spoofing involves transmitting fake signals that mimic those from GPS satellites, causing the GPS receivers to calculate incorrect Positioning, Navigation, and Timing (PNT) information. Recently, there has been a surge in GPS spoofing attacks targeting aircraft. Since GPS satellite signals are weak, the spoofed high-power signal can easily overpower them. These spoofed signals are often interpreted as valid by the GPS receiver, which can cause severe and cascading effects on air navigation. While much of the existing research on GPS spoofing focuses on technical aspects of detection and mitigation, human factors are often neglected, even though pilots are an integral part of aircraft operation and potentially vulnerable to deception. This research addresses this gap by conducting a detailed analysis of the behavior of student pilots when subjected to GPS spoofing using the Force Dynamics 401CR flight simulator with X-Plane 11 and a Cessna 172 equipped with Garmin G1000. Spoofing scenarios were implemented via custom scripts that altered navigational data without modifying the external visual environment. Thirty student pilots from the Embry-Riddle Aeronautical University Daytona Beach campus with diverse flying experience levels were recruited to participate in three spoofing scenarios. A pre-simulation questionnaire was distributed to measure pilot experience and confidence in GPS. In-flight decision-making during the spoofing attacks was observed, including reaction time to anomalies, visual attention to interface elements, and cognitive biases. A post-flight evaluation of workload was obtained using a modified NASA Task Load Index (TLX) method. This study provides a first step toward identifying human vulnerabilities to GPS spoofing amid the ongoing debate over GPS reliance.

## I. Introduction

The Global Positioning System (GPS) is the primary source of positioning, navigation, and timing (PNT) data for aircraft and air traffic management (ATM) [1]. However, GPS is vulnerable to spoofing attacks, where the deliberate transmission of counterfeit signals can lead receivers to compute incorrect PNT data [2]. The increasing number of spoofing attacks on aircraft has raised serious concerns in the aviation industry [3] because spoofing attacks can jeopardize flight safety if pilots receive and act on incorrect data and erroneous warnings about their aircraft's location [4]. GPS spoofing of aircraft can also lead to diplomatic incidents, evidenced by several politically motivated spoofing attacks resulting from tensions in Europe [5], [6]. Understanding and mitigating GPS spoofing attacks has thus become a critical priority to ensure the safety and resilience of the aviation industry.

Existing research has focused on the technical detection and mitigation of spoofing attacks, but the human factors aspects are often overlooked, despite pilots being central to navigation decisions and vulnerable to deception. Prior studies have explored the efficacy of spoofing detection methods (e.g., analysis of signal strength, origin, and quality, Machine Learning (ML) approaches, etc.) and mitigation efforts (e.g., multi-antenna techniques, suppression of fraudulent signals, and GPS data integrity monitoring) [7], [8], [9], [10], [11], [12]. Other research and development efforts have explored GPS enhancement systems, such as Receiver Autonomous Integrity Monitoring (RAIM) [13], which alerts pilots to GPS inconsistencies, and the Satellite-Based Augmentation System (SBAS) [14], which corrects GPS signal errors. The lack of research regarding the human element at the operational endpoint in a GPS spoofing attack remains a significant gap in the literature.

To address this gap, this study observed student pilot behavior during simulated GPS spoofing scenarios to examine their decision-making processes and reactions across varying levels of flight experience. This paper begins with a literature review to establish an understanding of spoofing attacks and factors that may affect pilots' ability to detect these attacks. The creation of three simulated GPS spoofing scenarios to emulate spoofing attacks is then described. Finally, a detailed analysis of the results is presented.

## II. Background and Related Work

It is first important to understand GPS, GPS spoofing, and factors that may influence pilots' reactions to spoofing attacks to contextualize the research approach.

### A. Global Positioning System (GPS)

The overall goal of GPS is to provide users with PNT services, which is made possible by collective information from at least four satellites [15]. The determination of a precise 3D location is based on the intersection of three spheres in space, each defined by the distance between the center of the satellite and the user's GPS receiver [16]. Finally, a fourth satellite is required to correct for the GPS receiver's internal clock error. Disrupting PNT signals was once limited to nation-states [17], but now, low-cost tools, such as HackRF (i.e., software defined radios), enable hackers to jam or spoof GPS. The structure of GPS is illustrated in Fig. 1.

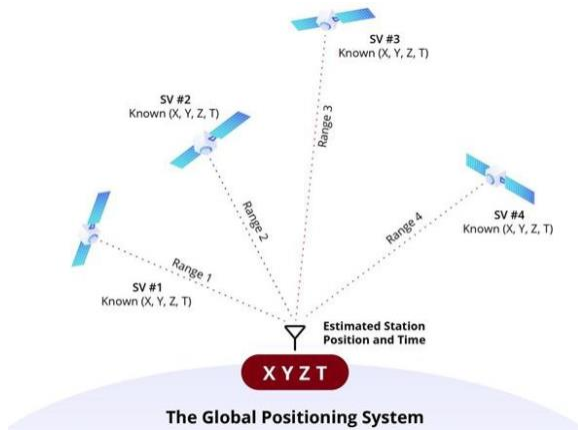

Fig. 1. At least four satellites are leveraged to determine the position in three dimensions (X, Y, Z) and GPS time (T) [15].

*B. GPS Spoofing*

The aim of GPS spoofing is to disrupt or corrupt PNT data, leading to erroneous positioning and navigation. This poses significant risks in aviation, including flight detours, GPS interference at airports, and falsified aircraft location [18]. There are three main types of spoofing attacks: (1) signal simulators, which imitate GPS signals; (2) spoofers based on a GPS receiver, which generate signals that synchronize with real signals at the right time and at the right power; and (3) highly advanced spoofers capable of perfectly synchronizing their signals with those of satellites [19].

On an aircraft, spoofing may be detectable through inconsistencies in flight parameters. However, if the GPS spoofing is very effective, then on-board safety and cyber-enhancement systems are unable to distinguish inconsistencies in PNT data, so no detection alarms or mitigation processes are triggered. These are worst-case scenarios, as the pilot then becomes the last remaining line of defense to detect a spoofing attack. Nevertheless, a pilot can still determine whether a GPS spoofing attack has occurred by examining inconsistencies between the predicted waypoint coordinates and those displayed, or discrepancies in estimated navigation or heading times. GPS spoofing is characterized by these and other ambiguous cues that often lead to a high degree of uncertainty, so it is crucial to understand what factors trigger a pilots' decision-making process if and when they recognize a GPS spoofing attack [20]. The extant literature has not thoroughly explored what factors influence a pilot's ability to detect these discrepancies, and because pilots are a significant safeguard in identifying whether a GPS spoofing attack has occurred, this study aims to respond to this gap.

*C. Pilots and GPS Spoofing*

Current cybersecurity efforts to mitigate GPS spoofing attacks on aircraft focus on technical approaches but neglect pilot recognition and response as possible obstacles or contributors to the severity of attacks. Pilot actions and decision-making are of particular importance in GPS spoofing attacks, as misinterpretations, low situational awareness, or overlooked indicators of interference may inadvertently contribute to the risks associated with GPS spoofing [21].

There are a number of factors which may affect a pilot's ability to detect a GPS spoofing attack. Pilots who have more flight experience may be more capable of identifying whether a GPS spoofing attack has taken place since empirical research has shown that experience affects reaction time, information processing, and subsequent decision-making [22], [23], [24]. Flight experience was therefore a factor in this study.

In addition, the speed with which the spoofing scenario is assessed may depend on the pilot's attention, which can be observed via eye movements and visual search patterns [25]. Eye tracking is essential for studying perception, decision-making, and cognitive load. In the absence of eye-tracking technology, eye tracking can be achieved by visually observing the eye movements of an individual and/or reviewing eye movements from a recording [26].

Moreover, previous research suggests that there is a significant lack of awareness among pilots regarding the vulnerability of aircraft to GPS spoofing attacks [15]. This lack of awareness may be compounded by the level of trust pilots have in GPS. Trust is an essential factor to consider in the context of GPS spoofing, since the more trust pilots have in their navigation system, the more they may not consider GPS spoofing as the cause of the perceived problem [27]. Trust in an attacked avionics system has previously been measured in [25] and [28] and was found to decrease after participants experienced a simulated attack. Thus, pilots may lack awareness to be vigilant against GPS spoofing attacks because of a baseline limited awareness of these attacks and high pre-existing trust in GPS.

Finally, GPS spoofing situations increase pilots' workload [29], which can impair their ability to safely operate aircraft or perform safety-related duties [30]. Workload was therefore measured using the NASA-Task Load Index (TLX) method to better understand pilot limitations in the face of GPS spoofing attacks.

### III. METHODOLOGY

The methodology for this study was organized into three phases: (1) <u>*a pre-simulation phase*</u>, which included a questionnaire to gather demographic information and assess student pilots' trust in GPS; (2) <u>*an in-flight simulation phase*</u>, involving observation of decision-making under simulated GPS spoofing scenarios (as illustrated in Fig. 2), with a focus on reaction time, coping strategies, and cognitive biases; and (3) <u>*a post-simulation phase*</u>, which evaluated perceived workload using a modified NASA-TLX questionnaire. All components of the study received approval from the Institutional Review Board (IRB) at Embry-Riddle Aeronautical University prior to participant recruitment. The following sections describe the three spoofing scenarios, followed by detailed descriptions of the pre-simulation, simulation, and post-simulation phases. The full set of simulation scripts and surveys can be found at https://github.com/ TAYLOK33/gps-spoofing-pilot-responses.

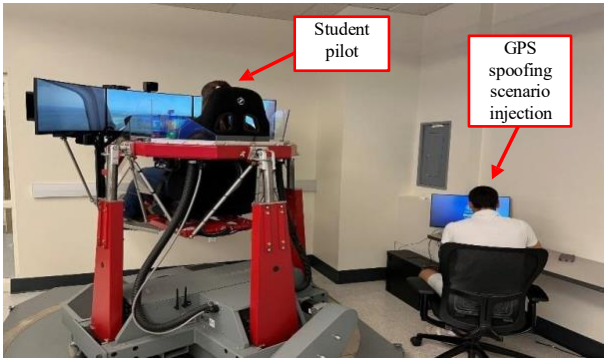

Fig. 2. Force Dynamics 401CR flight simulator with X-Plane 11 and a Cessna 172 equipped with Garmin G1000 used for GPS spoofing scenarios.

*A. GPS Spoofing Scenarios*

**Scenario 1**: In the first scenario, the flight plan was set from Naples Municipal Airport (KAPF) to Paige Field Airport in Fort Myers (KFMY). During the flight, GPS spoofing was introduced according to the parameters in Table I. If the student pilot failed to detect the spoofing, they were diverted from their intended trajectory of KFMY to Southwest Florida International Airport in Fort Myers (KRSW). This scenario was modeled after a real-life case [4], which was later reproduced in [20]. This scenario took approximately 15 minutes to complete. The objective of this scenario was to highlight the major impact that a falsification of distance, time, and heading measurement can have on aircraft navigation, potentially driving the pilot out of their flight plan and into unauthorized areas, which may be dangerous or prohibited.

TABLE I. SCENARIO 1 (FROM KAPF TO KFMY) PARAMETERS

| Flight Plan[a] | Without GPS Spoofing | GPS Spoofing Trigger | Consequences |
|---|---|---|---|
| 3 NM of KAPF-waypoint AVCOF | DIS = 6.6 NM; DTK = CRS = 062°; ETE = 4'34 | DIS = 2.0 NM | DIS and ETE are dropping more slowly to overtake the aircraft by 1.1 NM of AVCOF |
| AVCOF-KFMY | DIS = 24.8 NM; DTK = CRS = 348°; ETE = 14'29 | DIS = 23.5 NM | CRS = DTK delay, bringing the plane to the airport south of KRSW |
| AVCOF-KRSW | DIS = 20.2 NM; DTK = CRS = 357°; ETE = 11'29 | | ETE and DIS are reduced and run faster because the point is farther south |

[a.] NM = Nautical miles, DIS = Distance, DTK = Desired track, CRS = Course, ETE = Estimated time enroute, IAS = Indicated air speed, kT = knots (nautical miles per hour), AVCOF = X-Plane 11 waypoint name

In this scenario, if the pilot focused only on the primary flight display (PFD) data and blindly followed the GPS, they would have found themselves diverted from the original flight plan. On the other hand, if the pilot focused on the multi-function display (MFD), they should not have deviated from their original flight path. The actual path of the aircraft and its position were displayed on the MFD, and, given limitations of the flight simulator used in this study, the MFD information could not be spoofed. As a result, if the pilot had looked at the MFD and compared it to the PFD, the pilot would have been aware that they were deviating from the original trajectory. Based on which screen the pilot decided to focus on, it was possible to indirectly determine if the pilot had been able to diagnose the problem [29].

**Scenario 2**: The second scenario, 10 NM Approach Tampa International Airport (KTPA), falsified both the speedometer on the GPS PFD (by increasing the speed) and the heading of the trajectory. The heading initially planned to bring the aircraft to land on runway 19 right (19R). Instead, the heading was changed to deviate towards runway 19 left (19L). This scenario is detailed in Table II, and its duration was estimated to be 10 minutes.

TABLE II. SCENARIO 2 (KTPA 10 NM APPROACH SCENARIO) PARAMETERS

| Flight Plan | Without GPS Spoofing | GPS Spoofing Trigger | Consequences |
|---|---|---|---|
| 10 NM of KTPA-waypoint EGDUE | DIS = 4.5 NM; DTK = CRS = 181°; ETE = 3'30 | DIS = 3.5 NM | DIS and ETE drop faster. IAS increases to 130 kT |
| EGDUE[b]-KTPA 19R | DIS = 4.7 NM; DTK = CRS = 191°; ETE = 2'15 | DIS = 4.7 NM | CRS = DTK delay bringing the aircraft to runway 19L. IAS increases to 110 kT |

[b.] EGDUE = X-Plane 11 waypoint name

The objective of this scenario was to highlight the major impact of GPS spoofing during a risky phase of flight, namely the approach, during which the pilot's attention and workload are already very high. In addition, if the pilot deviated in this way, a spoofing attack of this nature could increase the risk of collision with other aircraft. In this scenario, weather conditions were intentionally set to deteriorated conditions so that pilots could not look out the window to reconcile their speed and heading with their outside surroundings. The expected reaction was that the pilot allowed themselves to deviate in the direction indicated by the falsified heading, which would have led them to the wrong runway. It was also expected that the pilot would become aware that the engine power was not increasing, and the gyroscopic anemometer indicated a lower speed than the PFD, which indicates GPS spoofing.

**Scenario 3:** Finally, the third scenario, Take-off from Daytona Beach International Airport (KDAB) to Ormond Beach Municipal Airport (KOMN), implemented gradual shifts of ground speed. This time, the speed decreased suddenly until it appeared as zero. The scenario is detailed in Table III, and its duration was estimated at 5 minutes. The objective of this scenario was to observe the pilot's information processing in the face of a more immediate unforeseen event.

TABLE III. SCENARIO 3 (TAKE-OFF SCENARIO FROM KDAB TO KOMN) PARAMETERS

| Flight Plan | Without GPS Spoofing | GPS Spoofing Trigger | Consequences |
|---|---|---|---|
| KDAB-KOMN | DIS = 8.0 NM; DTK = CRS = 343°; ETE = 6'42 | DIS = 6.0 NM | IAS will decrease to 60 kT and then to 0 |

During this scenario, the impact of GPS spoofing should have been discernible very quickly. The expected reaction was for the pilot to leverage their "System 1" impulsive thinking [31], which may have led them to incorrectly conclude that there was an engine problem when the airspeed suddenly dropped. If the pilot responded by increasing engine power, they may have noticed overspeed conditions, revealing a mismatch. The only reliable way for the pilot to discern that the problem was originating from a spoofing attack and not an engine problem was to visually confirm that the engine was functioning normally (i.e., by looking out the window); this would have engaged "System 2" thinking, which is characterized by logical reflection [31].

*B. Pre-Simulation*

The pre-simulation questionnaire obtained demographic information for each participant to understand their flight expertise, including their overall flight time, number of hours spent flying on the Cessna 172, and highest pilot license/certification. Thirty student pilots were recruited, and three levels of expertise were established, as shown in Table IV. "Beginners" (i.e., Level 1) corresponded to student pilots with a few hours of flight time and no license/certification. "Intermediate" (i.e., Level 2) corresponded to student pilots with a private pilot license (PPL) or a PPL with an instrument flight rules (IFR) rating. Finally, "Advanced" (i.e., Level 3), corresponded to students who had a commercial pilot license (CPL) or instructor ratings. In this study, a pilot's expertise was defined by flight hours and licensure rather than knowledge of GPS spoofing.

In addition to questions regarding flight experience, there were also five Likert-scale questions (on a scale of 1 to 6) that assessed participants' knowledge of GPS systems. These questions asked participants to rate their (1) GPS proficiency, (2) trust and (3) confidence in GPS, and (4) understanding of GPS and (5) its risks.

Before starting the simulation phase, all three flight plans were presented to the participants. Presenting this information was meant to mimic real-life flight situations, where pilots know their flight plan before taking off.

TABLE IV. STUDENT PILOT EXPERTISE LEVEL AND PARTICIPATION COUNT

| Level | Expertise | Participant Count |
|---|---|---|
| Level 1 | Beginners (few hours of flight time & no license/certification) | 7 (~23%) |
| Level 2 | Intermediate (PPL and/or IFR) | 14 (~47%) |
| Level 3 | Advanced (CPL and/or instructor ratings) | 9 (~30%) |
| | | Total = 30 |

*C. Simulation*

The researcher examined the reactions of the student pilots during the simulated GPS spoofing scenarios. First, the researcher visually observed the eye movements of pilots as they looked at the main interface elements represented by the following set: Aircraft, Navigation, and System Management (ANG), presented in Fig. 3. The aircraft (A), in cyan, was characterized by all the elements of the PFD, as well as the desired track (DTK) and estimated time enroute (ETE) on the MFD. The navigation set (N), shown in yellow in Fig. 3, included the MFD map. Finally, the management assembly (G), in red in Fig. 3, corresponded to the display indicating the status of the system (battery, engine, etc.) and all the primary instruments present on the instrument panel, whose operation was independent of the GPS. The objective of observing eye movements was to measure attention. Specifically, attention was defined as the amount of time (measured with a stopwatch) spent focusing on a single element of the ANG set, expressed as a proportion of the total gaze duration across all ANG elements. Eye tracking began once the GPS spoofing scenario was initiated.

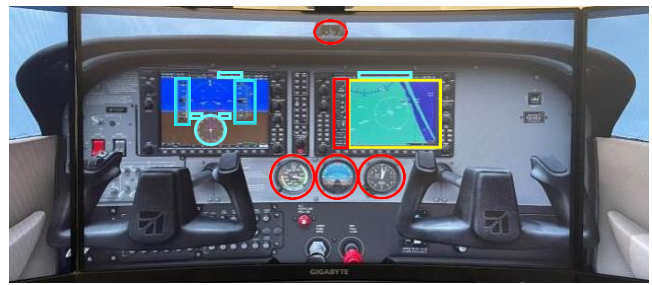

Fig. 3. Representation of the Aircraft (A) shown in cyan, Navigation (N) shown in yellow, and System Management (G) shown in red.

The participants' actions were also observed to understand potential biases, such as confirmation bias [32], anchoring bias [33], overconfidence bias [34], and focusing illusion bias [35], that may have influenced how they reacted to the spoofing scenarios. Behavioral observations and patterns, such as focusing predominantly on one ANG element or procedural errors that indicated the participant believed the problem was not spoofing-related (e.g., starting an engine failure procedure) were coded "yes" or "no" by a single analyst for each participant. The codes were designed to be provisional indicators of forms of cognitive bias, but they have not undergone construct validation. As a result, we treated these variables as exploratory descriptors rather than validated measures of bias.

*D. Post-Simulation*

The post-simulation questionnaire assessed the workload experienced by participants during the simulations. This was accomplished with a modified NASA-TLX survey. The NASA-TLX method is a subjective assessment tool that measures workload through six dimensions associated with a task: mental, physical, temporal load, performance, effort, and frustration [36]. This method, which is widely used in human factors research, was developed by the Human Performance Group at NASA.

For this study, the NASA-TLX questions were adapted to GPS spoofing scenarios. Each question was evaluated from 0 ("not at all") to 100 ("a lot"). The overall NASA-TLX score was calculated by averaging the scores divided by the six dimensions. The closer the score was to 100, the higher the perceived workload experienced by the pilots [37]. A correlation was conducted with the NASA-TLX results and pilot experience level.

## IV. RESULTS

### A. Confidence in GPS

The average confidence of pilots in GPS, based on the Likert-scale questions (on a scale of 1 to 6) in the pre-simulation survey, was 4.78/6, which is lower than the 5.59/6 reported by [25] for pilots under similar conditions (spoofing without warning). This difference may stem from the higher expertise of the sample in [25], composed exclusively of professional pilots holding at least a multi-crew pilot license (MPL), which was equivalent to Level 3 in this study.

When analyzing GPS confidence by pilot level, the results show that Level 3 pilots displayed higher confidence. This is shown by the results of a between-subjects analysis of variance (ANOVA). The test result ($F = 3.23 > F_{critical} = 2.51$, $p = 0.0551$) was not significant at the 5% threshold, but it was significant at the 10% threshold. Since this was an exploratory study, the 10% threshold was utilized, indicating that expertise marginally influenced GPS confidence. Experienced pilots tended to trust their GPS more. Moreover, a Pearson correlation analysis indicated that the most influential factor on GPS confidence was risk awareness, with a strong positive correlation ($r = 0.707$). This supports findings from [27] and [37], emphasizing that knowledge-based trust is more robust than passive or indirect trust.

### B. Eye Tracking and Attention

Analysis of pilots' eye movements across cockpit elements A, N, and G (see Fig. 3) revealed that A elements captured the most attention, with an average of 70.7% across all three spoofing scenarios, as seen in Fig. 4. This aligns with the results of [25], which found a primary focus on A elements (59.68%) in spoofing scenarios without warning. However, coefficients of variation (CV) for attention on A, N, and G are high, indicating strong variability, which may be due to pilot expertise. For example, CVs for G in Scenarios 1 and 3 in Fig. 4 exceed 145%, reflecting large dispersion in focus patterns.

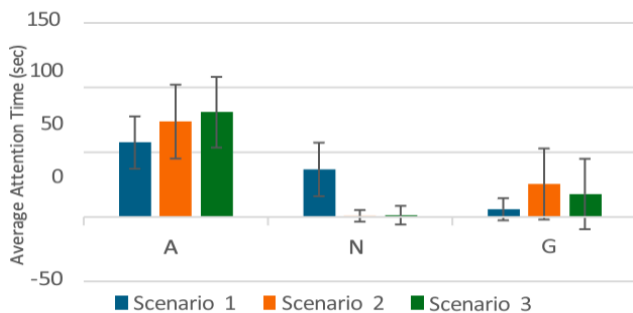

Fig. 4. Average distribution of pilot attention times on A, N, and G elements in the three spoofing scenarios.

In Scenario 1, there was a substantial change in average attention time between the three elements, with A receiving the most attention. In Scenarios 2 and 3, detecting spoofing required confronting A and G elements, which may explain why N received almost no attention in these scenarios. A linear regression indicated a significant positive correlation between time to detect anomalies and attention to A elements ($p = 0.041$, $r = 0.189$), suggesting that excessive attention to instruments can delay anomaly detection.

Anomaly response times between the three levels of pilot expertise were also compared, as shown in Table V. A one-way, between-subjects ANOVA indicated that expertise had a significant impact on reaction time, $F = 221.90 > F_{critical} = 7.71$, $p < 0.001$. This suggests that pilot expertise directly influences attention allocation and reaction efficiency. Experienced pilots scanned more evenly, confronting multiple data sources, which may have enhanced their ability to detect the GPS spoofing faster.

TABLE V. AVERAGE GPS SPOOFING RESPONSE TIME ACROSS THREE LEVELS OF EXPERTISE

| Pilot Expertise Level | Average GPS Spoofing Response |
|---|---|
| Level 1 | 52.29 sec |
| Level 2 | 46.93 sec |
| Level 3 | 42.00 sec |
| | Overall average = 46.7 sec |

### C. Pilot Reactions and Decision-Making Processes

Pilot reactions observed across the three scenarios allowed for an overall conclusion to be drawn for each case, indicating whether the pilot successfully counteracted the effects of GPS spoofing. Each scenario was evaluated based on three outcome categories:

*1) Success*: The pilot correctly interpreted the situation and/or was not deceived by the GPS spoofing.

*2) Failure*: The pilot was misled by the spoofing. In Scenario 1, this meant completely deviating from the intended flight path. In Scenarios 2 and 3, this meant misinterpreting speed anomalies, believing the airspeed indicator had failed, and initiating incorrect procedures while ignoring correct data shown by the gyroscopic airspeed indicator or the RPM gauge.

*3) Danger*: The pilot lost control of the situation, leading to hazardous conditions such as an approach that was dangerously low, a complete reduction in speed risking a stall, or a steep dive to regain airspeed.

From these observations, the number of outcomes categorized as "Failure" and "Danger" were tallied per pilot over the three flight plan scenarios, segmented by pilot expertise level. A Chi-squared ($\chi^2$) test was conducted to evaluate the association between pilot experience level and outcome. The resulting *p*-value was 0.00484, indicating a statistically significant association between pilot experience level and how effectively they managed GPS spoofing situations.

Although all participants appeared to initially assume that radio/navigation instrument failure was the cause of problems in each scenario, experienced pilots were more likely to initiate radio contact when uncertain. This behavior suggests that experience contributes to improved mental workload management and cognitive flexibility, enabling better responses in ambiguous situations. Despite this, all expertise levels showed a general lack of awareness of GPS spoofing, leading to misinterpretation of the actual issue.

This was further confirmed through the observation of cognitive biases observed during the scenarios. The bar chart in Fig. 5 highlights that confirmation bias was the most prevalent, indicating pilots tended to seek information that confirmed their initial (often incorrect) hypotheses about the nature of the problems in each scenario. Further analysis showed that confirmation bias was uniformly distributed across all expertise levels, as evidenced by a Chi-squared test ($\chi^2 = 0.16$; $df = 2$; $p = 0.92$). This suggests that this particular bias is not mitigated by experience, possibly due to its deeply rooted nature in human cognition. In uncertain situations lacking clear cues, even experienced pilots may fall back on biased reasoning to maintain internal consistency.

Conversely, other biases were significantly correlated with expertise level.

*1) Overconfidence Bias ($\chi^2 = 20.26$; $p < 0.001$):* A Chi-squared test of independence showed that overconfidence bias differed significantly between the three levels of pilot experience. Novice pilots frequently overestimated the reliability of the GPS despite visible inconsistencies, while experts were more cautious. Interestingly, this occurred despite student pilots reporting lower confidence levels, highlighting a discrepancy between perceived and actual behavior. Training that encourages critical assessment of navigation systems could mitigate this bias.

*2) Focus Illusion Bias ($\chi^2 = 17.97$; $p < 0.001$):* A Chi-squared test of independence showed that focus illusion bias differed significantly between the three levels of pilot experience. This bias primarily affected less experienced pilots, who tended to concentrate excessively on the GPS at the expense of other information sources and gave disproportionate attention to A elements. Early training should emphasize instrument cross-checking strategies.

*3) Anchoring Bias ($\chi^2 = 11.35$; $p = 0.005$):* A Chi-squared test of independence showed that anchoring bias differed significantly between the three levels of pilot experience. While less pronounced, this bias also declined with experience. Novices often clung to their initial assumptions (e.g., the flight plan) despite contradictory indicators.

The results suggest that pilot reactions and decision-making processes were influenced by their level of expertise, with overconfidence, focus illusion, and anchoring biases being significantly reduced through experience. However, other biases, such as confirmation bias and the misinterpretation of error nature, appear more universal.

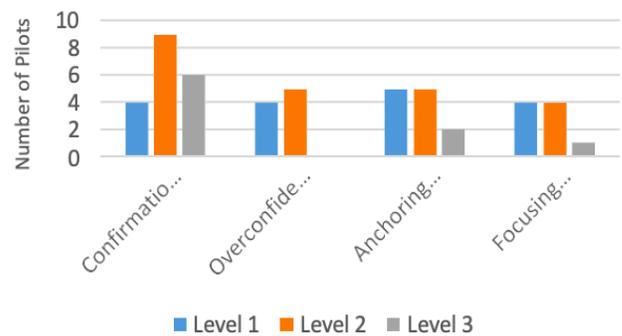

Fig. 5. Cognitive bias in pilots during scenarios according to their level of expertise.

Additionally, statistical analysis of NASA-TLX workload assessments showed no significant effect of pilot expertise on perceived workload, shown in Fig. 6, which illustrates the distribution of workload scores by experience level.

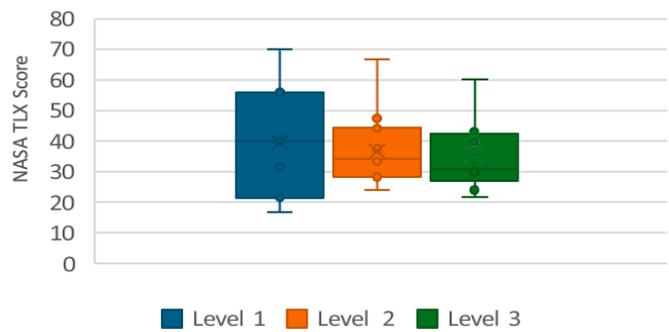

Fig. 6. NASA-TLX test results by pilot expertise level.

V. DISCUSSION

This study constitutes a preliminary step toward a more comprehensive analysis of pilot reactions to GPS spoofing. The sample size of 30 pilots was limited in scope, particularly due to the unequal distribution among the different experience levels. Future studies should recruit a broader, more diverse population of experienced pilots and work toward construct validity for measuring confidence in GPS. Expanding upon this research with pilots and Air Traffic Control (ATC) systems integrated into the study methodology would also provide a more robust and comprehensive understanding of pilot responses in real-world spoofing environments.

The available tools did not allow for a complete spoofing of the GPS system. For instance, it was not possible to alter the MFD displays, such as the aircraft's position and trajectory line, which inevitably influenced pilot decisions and limited the number of operational errors observed. Funding and resource limitations also prevented the researchers from obtaining professional-grade eye-tracking equipment. Future work should employ more robust eye-tracking methods.

Additionally, given that flight safety stakes are significantly lower in a simulator compared to a real Cessna 172, some pilots exhibited behaviors that would be considered significant safety hazards in an actual flight. This context may also explain the relatively low perceived workload, which would likely be higher under real-world conditions.

The evaluation of cognitive biases in this study are hypothesis-generating. Without using validated measures of the biases in this study, our observations do not establish the presence or extent of cognitive bias, nor do they indicate causation of any kind. However, our observations do act as exploratory descriptors of bias. Future work should employ validated bias measurement tools, perhaps incorporating a panel of subject-matter experts and interrater reliability measures to assess validity of these bias indicators.

Moreover, the experimental setup deliberately excluded all GPS backup or reinforcement systems. This choice was made to focus exclusively on human factors, using highly sophisticated spoofing attacks that did not trigger any alarms. While this isolation aids in analyzing human behavior, it is not realistic to assume that safety depends solely on the pilot. Alarm systems would likely have supported better diagnosis, and the air traffic control (ATC) role was voluntarily omitted despite its crucial importance. In reality, ATC could detect an aircraft's deviation (assuming their systems are not also spoofed) and either assist or alert the pilot accordingly.

The results from this study have practical implications for pilot training. Previous research on the limitations of pilot training have emphasized the importance of including GPS spoofing in training programs [25]. This may result in greater emphasis on training pilots in alternative systems that do not require the use of GPS, such as inertial navigation systems (INS/IRS) [38].

## VI. CONCLUSION

This study aimed to investigate how student pilots responded to GPS spoofing attacks within a flight simulator environment. To model these attacks, three spoofing scenarios were developed in which key navigational data, including heading and speed, were falsified. The objective was to observe pilots' decision-making processes and reactions across varying levels of experience.

In this investigation, the initial assumptions were grounded in previous research, which highlighted a lack of awareness among pilots regarding the risks associated with GPS spoofing. This deficiency was believed to be linked to a high level of trust in the GPS system and a general lack of skepticism and absence of practical training addressing GPS spoofing in current pilot training programs. It was thus hypothesized that pilot expertise would play a significant role, as suggested by [5], particularly in anomaly detection, decision-making, and reaction processes. Specifically, it was expected that novice pilots would be more likely to over-rely on GPS data, failing to question its reliability and remaining focused on irrelevant interface elements during abnormal scenarios.

Conversely, experienced pilots, being more aware of the system's vulnerabilities, were expected to detect inconsistencies more rapidly and revert to alternative navigation methods. The findings of this observational study indicate the pilot experience marginally affects confidence in GPS. This highlights the necessity of raising awareness among novice pilots about the associated with GPS and the advantages of modern, alternative navigation tools.

Although experienced pilots generally demonstrated faster reaction times in detecting scenario anomalies, reaction times did not always translate into an accurate diagnosis of spoofing. Error analysis and the identification of possible cognitive biases revealed a general lack of specific knowledge about GPS spoofing across all experience levels. This suggests the need to include dedicated training modules focused on managing uncertainty and fostering critical thinking in atypical scenarios. Surprisingly, one counterintuitive finding emerged: Overconfidence bias was primarily observed among pilots in Levels 1 and 2, despite their GPS confidence scores not being significantly higher than those of other groups.

The study also highlighted the major challenges that pilots face when detecting a spoofing attack. This insight leads us to recommend raising awareness, updating training protocols, and improving flight preparation procedures for GPS spoofing. These recommendations seek to make pilots more aware of GPS vulnerabilities and enhance their ability to respond appropriately in the face of such threats.